\documentclass[aps,prl,reprint,superscriptaddress,nofootinbib,preprintnumbers]{revtex4-2}

\usepackage{listings}
\usepackage{xspace}
\usepackage{slashed}
\usepackage[caption=false]{subfig}
\usepackage{graphicx}
\usepackage[colorlinks=true,linkcolor=blue,citecolor=blue,urlcolor=blue]{hyperref}
\usepackage{xcolor}
\usepackage{amsmath,amssymb,bm}
\usepackage[most]{tcolorbox}
\tcbuselibrary{breakable,skins}

\newtcolorbox{promptblocktex}[1][]{
  enhanced,
  breakable,
  colback=black!3,
  colframe=black!35,
  colbacktitle=black!18,
  coltitle=black,
  boxrule=0.45pt,
  arc=2pt,
  left=6pt,
  right=6pt,
  top=6pt,
  bottom=6pt,
  title={#1},
  fonttitle=\ttfamily\footnotesize\bfseries,
  fontupper=\small,
  before skip=8pt,
  after skip=8pt
}

\newcommand{\collideranget}{\textsc{ColliderAgent}\xspace}
\newcommand{\magnus}{\textsc{Magnus}\xspace}
\newcommand{\madgraph}{\textsc{MadGraph}\xspace}
\newcommand{\madanalysis}{\textsc{MadAnalysis}\xspace}
\newcommand{\mathematica}{\textsc{Mathematica}\xspace}
\newcommand{\feynrules}{\textsc{FeynRules}\xspace}
\newcommand{\pythia}{\textsc{Pythia}\xspace}
\newcommand{\delphes}{\textsc{Delphes}\xspace}

\newcommand{\PKUPhys}{School of Physics, Peking University, Beijing 100871, China}
\newcommand{\PKUHEP}{Center for High Energy Physics, Peking University, Beijing 100871, China}
\newcommand{\CCNU}{Institute of Particle Physics and Key Laboratory of Quark and Lepton Physics (MOE), Central China Normal University, Wuhan, Hubei 430079, China}
\newcommand{\CSRC}{Beijing Computational Science Research Center, Beijing 100193, China}
\newcommand{\ZZU}{School of Physics, Zhengzhou University, Zhengzhou 450001, China}
\newcommand{\PKUNPT}{State Key Laboratory of Nuclear Physics and Technology, Peking University, Beijing 100871, China}

\begin{document}

\title{An End-to-end Architecture for Collider Physics and Beyond}

\author{Shi Qiu}
\affiliation{\PKUPhys}

\author{Zeyu Cai}
\affiliation{\PKUPhys}

\author{Jiashen Wei}
\affiliation{\PKUPhys}

\author{Zeyu Li}
\affiliation{\PKUPhys}
\affiliation{\PKUHEP}

\author{Yixuan Yin}
\affiliation{\PKUPhys}

\author{Qing-Hong Cao}
\affiliation{\PKUPhys}
\affiliation{\ZZU}
\affiliation{\PKUHEP}

\author{Chang Liu}
\affiliation{\PKUPhys}
\affiliation{\PKUNPT}

\author{Ming-xing Luo}
\affiliation{\CSRC}

\author{Xing-Bo Yuan}
\affiliation{\CCNU}

\author{Hua Xing Zhu}
\affiliation{\PKUPhys}
\affiliation{\PKUHEP}

\preprint{CPTNP-2026-012}

\begin{abstract}
We present, to our knowledge, the first language-driven agent system capable of executing end-to-end collider phenomenology tasks, instantiated within a decoupled, domain-agnostic architecture for autonomous High-Energy Physics phenomenology. Guided only by natural-language prompts supplemented with standard physics notation, \textsc{ColliderAgent} carries out workflows from a theoretical Lagrangian to final phenomenological outputs without relying on package-specific code. In this framework, a hierarchical multi-agent reasoning layer is coupled to \textsc{Magnus}, a unified execution backend for phenomenological calculations and simulation toolchains. We validate the system on representative literature reproductions spanning leptoquark and axion-like-particle scenarios, higher-dimensional effective operators, parton-level and detector-level analyses, and large-scale parameter scans leading to exclusion limits. These results point to a route toward more automated, scalable, and reproducible research in collider physics, cosmology, and physics more broadly.
\end{abstract}

\maketitle

A central task of High-Energy Physics (HEP) phenomenology is to identify the fundamental constituents of matter and their interactions from first principles. In practice, especially
in searches for physics beyond the Standard Model (BSM), this requires translating theoretical concepts, most often encoded in a Quantum Field Theory (QFT) Lagrangian, into physical observables that can be confronted with experiment.

Over the past decades, phenomenological studies of Physics beyond the Standard Model (BSM) have led to two major assets of the field. The first is a mature research workflow, built from accumulated experience in turning theoretical models into quantitative predictions~\cite{Christensen:2009jx}. The second is a rich software ecosystem, in which specialized packages such as \feynrules \cite{Alloul:2013bka}, \textsc{FeynArts} \cite{Hahn:2000kx}, \textsc{FeynCalc} \cite{Shtabovenko:2016sxi, Shtabovenko:2020gxv,Shtabovenko:2023idz}, \madgraph \cite{Alwall:2014hca}, \pythia \cite{Bierlich:2022pfr}, and \delphes \cite{deFavereau:2013fsa}\ carry out individual stages of this workflow. Executing phenomenological studies by orchestrating such tools has consequently become a standard research paradigm in HEP~\cite{Boyle:2022cvo}. Although this workflow has enabled much of modern HEP phenomenology, the underlying tools differ substantially in syntax and usage conventions. The manual orchestration of heterogeneous toolchains therefore remains a major bottleneck in the field.

Although existing automation frameworks~\cite{Porod:2003um,Staub:2008uz,Dercks:2016npn} can streamline selected stages of these workflows, they typically remain limited to restricted classes of phenomenological problems. Recent agent-based studies have also begun to address selected components of HEP workflows~\cite{Menzo:2025cim,Plehn:2026gxv,Esmail:2026jpb,Badea:2026klb}, but not the full chain from a theoretical Lagrangian to final phenomenological outputs. Fully autonomous end-to-end HEP phenomenology nevertheless remains an open challenge. In this work, we propose a decoupled, domain-agnostic architecture for autonomous HEP phenomenology, and instantiate it for collider studies as \collideranget. The framework separates cognitive reasoning from analytical and numerical execution: specialized sub-agents translate natural-language instructions and standard physics notation into tool-specific operations, while \magnus~\cite{Magnus} provides a unified execution backend for phenomenological calculations and simulation toolchains.

To our knowledge, \collideranget\ is the first language-driven agent system to execute end-to-end collider-phenomenology tasks, from a theoretical Lagrangian to final phenomenological outputs. We assess its performance on representative benchmark reproductions spanning parton-level and detector-level analyses, specific BSM scenarios and Effective Field Theory (EFT), and tasks ranging from differential kinematic distributions to exclusions in model parameter space, all specified through natural-language prompts supplemented only by conventional physics expressions and numerical inputs. Together, these results point to a concrete route toward more automated, scalable, and reproducible phenomenological research, with potential applications extending from current LHC studies to future collider programs such as the CEPC, FCC, and Muon Collider.

\vspace{1.5em}

\begin{figure*}[t]
\centering
\includegraphics[width=\linewidth]{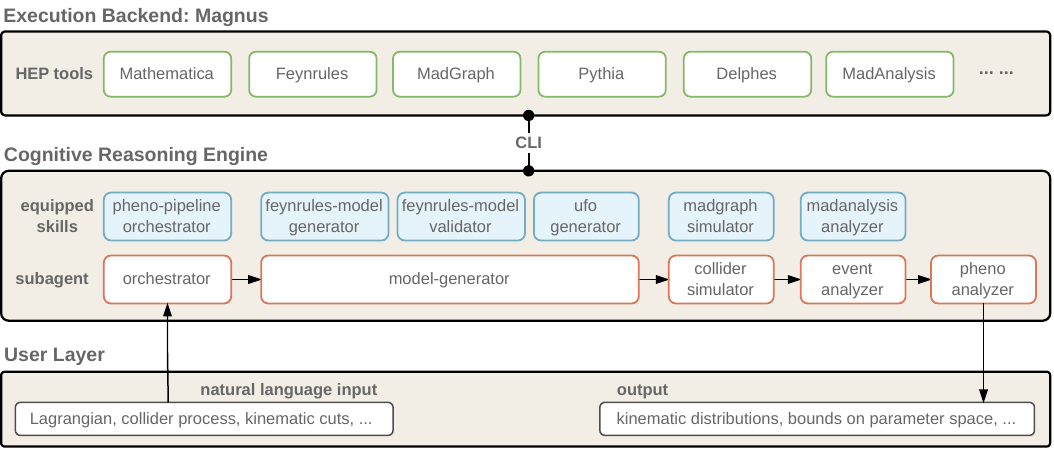}
\caption{Decoupled architecture of \collideranget. A hierarchical cognitive reasoning engine assigns phenomenological tasks to specialized sub-agents equipped with dedicated \textsc{Agent Skills}, which communicate through a standardized CLI with \magnus, the unified execution backend hosting the underlying HEP tools. The framework translates natural-language input and standard physics notation into final phenomenological outputs such as kinematic distributions and exclusion limits.}
\label{fig:architecture}
\end{figure*}

\noindent\textbf{Autonomous Framework.} The aim of autonomous collider phenomenology is to translate theoretical concepts, specified through natural-language instructions and standard physics notation, including \LaTeX\ expressions, into physical observables and model-dependent phenomenological inferences, ranging from kinematic distributions to exclusion limits in parameter space. To this end, we propose a decoupled architecture that is domain-agnostic in design, separating cognitive reasoning from analytical and numerical computation. As shown in Fig.~\ref{fig:architecture}, a hierarchical multi-agent cognitive layer assigns phenomenological tasks to specialized sub-agents, which interface through a standardized Command Line Interface (CLI) with \magnus, a general-purpose execution backend for phenomenological calculations and simulation toolchains. This reasoning-execution separation is designed to apply more broadly across HEP phenomenology. In this work, we instantiate this architecture as \collideranget. This section focuses on three elements: specialized cognitive delegation, a unified execution environment, and closed-loop validation and error correction.

\paragraph{Specialized Sub-agents.}
Collider phenomenology relies on a heterogeneous software ecosystem whose syntactic and operational conventions differ substantially across tools, from particle naming schemes to the built-in functions used to define models and implement analyses. A central observation underlying our design is that, despite this heterogeneity, many HEP tools are controlled through lightweight steering scripts or configuration files, such as the \texttt{.fr} model files used in \feynrules. Our cognitive layer exploits this shared structure through a hierarchical design centered on a master orchestrator. Depending on the user request, the orchestrator can either assemble an end-to-end workflow, from Lagrangian specification to event analysis, or invoke individual capabilities and assign the corresponding tasks to specialized sub-agents. Each sub-agent operates with a dedicated \textsc{Agent Skills}~\cite{AgentSkills}, namely a portable task-specific instruction module coupled to a reference card based on official software documentation~\cite{Alloul:2013bka,Alwall:2014hca,Conte:2012fm}, systematically curated by the authors on the basis of domain expertise and refined through extensive iterative testing. This design gives each sub-agent a complete local context for its assigned task while preserving portability across agent frameworks. To coordinate information across sub-agents, the system also maintains structured intermediate progress records that capture the essential physical and procedural state in a compact form, allowing later stages to recover the necessary context without carrying the full preceding history. Together, these design choices allow the system to translate high-level physical objectives into tool-specific operations while keeping the context passed across sub-agents compact.

\paragraph{Execution Backend.}
A practical obstacle in collider phenomenology is the difficulty of maintaining a consistent execution environment across heterogeneous software packages with nontrivial dependencies. In our framework, the cognitive sub-agents are separated from execution and dispatch all computational tasks to \magnus, a unified execution backend for phenomenological calculations that integrates the standard toolchain, including \mathematica\ (\textsc{Wolfram Engine}), \feynrules, \madgraph, \pythia, \delphes, and \madanalysis. By providing a preconfigured software environment, \magnus reduces setup overhead and improves reproducibility across runs. It is accessed through the CLI and can be deployed either locally, for example through Docker containers, or on high-performance computing clusters through workload managers such as \textsc{Slurm}. This separation allows the reasoning layer to remain largely independent of the underlying execution environment while supporting both interactive studies and large-scale analytical or numerical workloads, and is therefore consistent with the broader domain-agnostic design of the framework.

\begin{figure*}[t] 
    \centering
    \subfloat{\includegraphics[width=0.48\textwidth]{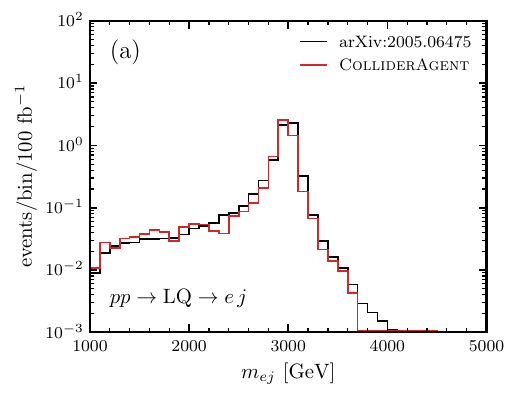}}
    \hfill
    \subfloat{\includegraphics[width=0.48\textwidth]{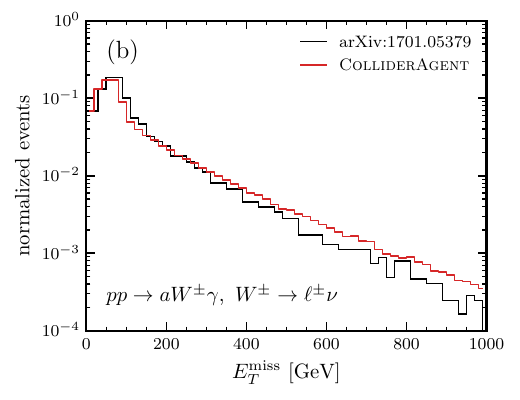}}

    \subfloat{\includegraphics[width=0.48\textwidth]{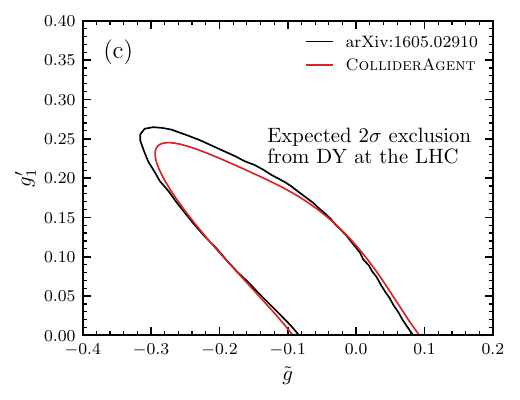}}
    \hfill
    \subfloat{\includegraphics[width=0.48\textwidth]{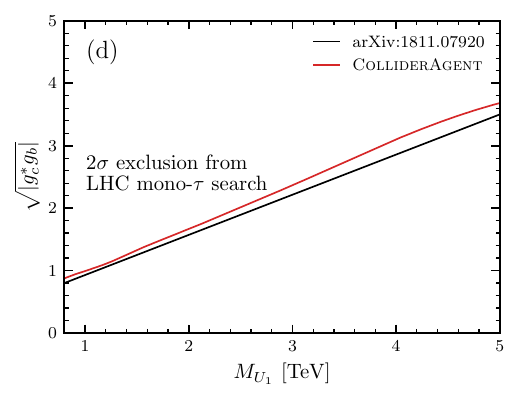}}
    \caption{Representative literature reproductions by \collideranget. (a) $m_{ej}$ distribution for $pp \to \text{LQ} \to e \, j$ at the LHC in the minimal scalar LQ model. (b) Normalized missing transverse energy distribution for $pp \to a W^\pm \gamma$ with $W^\pm \to \ell^\pm \nu$ at the LHC in the ALP EFT. (c) Expected $2\,\sigma$ exclusion contours for a $Z^\prime$ model from Drell-Yan production at the LHC. (d) $2\,\sigma$ exclusion contour for the $U_1$ LQ from the LHC mono-$\tau$ search. Panels (a)-(d) reproduce Refs.~\cite{Buonocore:2020erb,Brivio:2017ije,Accomando:2016sge,Greljo:2018tzh}, respectively. See text for details.}
    \label{fig:reproduction}
\end{figure*}

\paragraph{Validation and Self-Correction.}
Autonomous execution requires the ability to detect and resolve common failures without manual intervention. To this end, the framework applies a validation-and-repair loop before launching large-scale numerical calculations and simulations. In particular, the \texttt{model-generator} sub-agent invokes a dedicated \texttt{model-validator} \textsc{Skills} to inspect the generated model at several levels, including syntax checks, built-in \feynrules\ consistency tests such as Hermiticity, and test loading of the exported UFO model in \madgraph. If a check fails, for example because of a missing Hermitian conjugate in the generated \feynrules\ model file or a UFO syntax error, the \texttt{model-generator} parses the diagnostic output, revises the model, and repeats the validation step. This procedure reduces manual debugging and helps ensure that common model-construction errors are caught before computationally intensive phenomenological calculations and simulations are performed.

Additional technical details of the framework implementation, including the agent hierarchy and the \magnus execution environment, are provided in the Supplemental Material, together with usage examples and step-by-step instructions.

\vspace{1.5em}

\noindent\textbf{Physics Validation.} To assess the feasibility, robustness, efficiency, and scalability of \textsc{ColliderAgent}, we benchmark it on a diverse set of collider-phenomenology tasks, ranging from parton-level to detector-level analyses, from specific new-physics scenarios to higher-dimensional EFT operators, from differential kinematic distributions to exclusions in model parameter space, and from LHC to Muon Collider studies. Figure~\ref{fig:reproduction} summarizes four representative examples discussed in the main text, while additional benchmark reproductions and the corresponding prompts for all benchmark tasks are presented in the Supplemental Material. All benchmarks are specified through natural-language prompts, supplemented only by standard physics expressions and numerical inputs written in conventional notation, without package-specific code or executable scripts.

We begin with resonant single scalar-leptoquark production at the LHC, following Ref.~\cite{Buonocore:2020erb}. The prompt specified the minimal scalar leptoquark Lagrangian in Refs.~\cite{Bauer:2015knc,Schmaltz:2018nls}, the signal process $pp\to \mathrm{LQ}\to e\,j$, and the event selection and histogram prescription for the $m_{ej}$ distribution. This benchmark is nonstandard because the resonance is produced through lepton-quark collisions, requiring the correct use of the \texttt{LUXlep} PDF set~\cite{Manohar:2016nzj,Manohar:2017eqh,Buonocore:2020nai} to account for leptons inside the proton. The prompt further specified the technical workaround needed for showering: since \pythia\ cannot handle incoming leptonic partons, the initial-state leptons in the \texttt{LHE} events were to be replaced by photons before showering in \pythia. The agent then carried the workflow through showering, detector simulation with \delphes, and the corresponding event analysis, with complexity comparable to \madanalysis\ expert mode. As shown in Fig.~\ref{fig:reproduction}(a), the agent successfully handled these technical subtleties, completed the full multi-stage workflow, and reproduced the characteristic $m_{ej}$ resonance peak of single-LQ production. This example shows that the framework can reliably execute a broad class of technically demanding end-to-end collider-simulation workflows, including those relevant to state-of-the-art collider phenomenology.

Moving beyond canonical resonance searches, we next tested the framework in an axion-like particle (ALP) EFT benchmark. The prompt specified the two bosonic operators $\mathcal{A}_{\tilde{W}}$ and $\mathcal{A}_{\tilde{B}}$ in Refs.~\cite{Georgi:1986df,Brivio:2017ije}, with Wilson coefficients related by $c_{\tilde{B}}=-\tan^2\theta_W\,c_{\tilde{W}}$ so that $g_{a\gamma\gamma}=0$, leaving $c_{\tilde{W}}$ as the only independent coupling. It also provided the parton-level setup and event-selection cuts, and tasked the agent with reproducing the normalized $\slashed{E}_T$ distribution for $pp \to a W^\pm \gamma$ with $W^\pm \to \ell^\pm \nu$ at the $13~\mathrm{TeV}$ LHC. This benchmark probes the agent's ability to generate the correct EFT model files for the electroweak vertices induced by higher-dimensional operators, handle a three-body final state, and reconstruct invisible-particle kinematics. As shown in Fig.~\ref{fig:reproduction}(b), the agent successfully reproduces the normalized $\slashed{E}_T$ spectrum reported in Ref.~\cite{Brivio:2017ije}, including the characteristically hard tail induced by the $\mathcal{A}_{\tilde{W}}$ operator. This demonstrates that the framework can faithfully translate an EFT specification into the corresponding model files, matrix elements, and differential collider observables.

To demonstrate the framework's efficiency and scalability, we instructed the system to perform an exhaustive 2D parameter-space scan for a $U(1)^\prime$ extension of the SM in Refs.~\cite{Basso:2008iv,Coriano:2015sea}. The prompt given to the agent specified the final objective of deriving the expected LHC constraints on the $U(1)^\prime$ gauge coupling $g_1^\prime$ and the gauge-mixing parameter $\tilde{g}$ for different $m_{Z^\prime}$ hypotheses, together with the intermediate steps and essential physics inputs required to achieve this goal. These included the $Z^\prime$-fermion interaction Lagrangian, the signal process $pp \to \ell^+\ell^-$ mediated by the $Z^\prime$, the SM Drell-Yan background, and the definition of the statistical significance, which together provide the inputs needed to set the limits. Although the underlying physics is well understood, translating the Lagrangian into a complete parameter-space scan remains technically cumbersome. After generating the correct \feynrules model file, the agent used the \textsc{Magnus} platform to compute the relevant cross sections and automatically determined and executed the full parameter-space scan. As shown in Fig.~\ref{fig:reproduction}(c), it successfully reproduced the 95\% CL exclusion contours on the $(\tilde{g}, g'_1)$ plane for $M_{Z'} = 2, 2.5$ and $3~\mathrm{TeV}$, in agreement with Ref.~\cite{Accomando:2016sge}. This example illustrates the ability of the system to compress weeks of manual scripting, job orchestration, and data aggregation into a fully autonomous workflow completed within hours.

As a final and more stringent benchmark, we challenged \textsc{ColliderAgent} with a detector-level reproduction of the $U_1$ vector-leptoquark analysis in Ref.~\cite{Greljo:2018tzh}, targeting the mono-$\tau$ signature at the LHC~\cite{ATLAS:2018ihk,CMS:2018fza}. The prompt set the goal of deriving the $2\sigma$ exclusion contour in the $(\sqrt{|g_c g_b|}, M_{U_1})$ plane and provided the key ingredients needed for the analysis. Here, $g_c$ and $g_b$ denote the $U_1$ couplings to the charged currents involving the $c$ and $b$ quarks, respectively. On this basis, the agent autonomously reconstructed the full workflow, including event generation for the signal process $pp \to \tau \nu$, showering and hadronization, detector simulation with ATLAS and CMS configurations, experiment-specific event selection, and the extraction of the exclusion contour by comparing the resulting signal templates with published LHC data in a profile-likelihood analysis. This task required the seamless coordination of a heterogeneous toolchain, from the agent's generation of the corresponding \feynrules and UFO model files to \madgraph, \pythia, \delphes, and \madanalysis, without human intervention. As shown in Fig.~\ref{fig:reproduction}(d), the agent successfully reproduced the exclusion contour reported in Ref.~\cite{Greljo:2018tzh}. This example demonstrates that the system can operate not only on parton-level phenomenology and parameter scans, but also across the full detector-level inference pipeline required in realistic collider studies.

\vspace{1.5em}

\noindent\textbf{Conclusion.} We have presented a decoupled, domain-agnostic architecture for autonomous high-energy-physics phenomenology, and instantiated it in this work as \collideranget\ for collider phenomenology. Combining specialized sub-agents, a unified execution backend (\magnus), and a validation-and-repair loop, the framework translates natural-language instructions and standard physics notation into executable phenomenological workflows without relying on package-specific code. To our knowledge, \collideranget\ is the first language-driven agent system to execute end-to-end collider-phenomenology tasks, from a theoretical Lagrangian to final phenomenological outputs.

Across representative benchmarks, \collideranget\ reproduces literature results spanning parton-level and detector-level analyses, specific new-physics scenarios and higher-dimensional EFT operators, and tasks ranging from differential kinematic distributions to exclusions in model parameter space. These results show that language-driven agents can perform technically demanding collider-phenomenology studies while remaining grounded in established physics toolchains.

The same framework is also expected to facilitate phenomenological studies for future collider programs, including the CEPC, FCC and Muon Collider. Related agent-based efforts are also emerging in experimental workflows~\cite{Badea:2026klb,DrSai}, pointing to a broader role for autonomous AI systems across HEP. More broadly, this work suggests a route toward more automated, scalable, and reproducible research across collider physics, cosmology and physics in general.

\vspace{1.5em}

\begin{acknowledgments}

We thank the authors of all HEP packages used in this work for making their tools publicly available. This work is supported by the National Natural Science Foundation of China under contract No.~12425505, 12135006, 12575099, 12235001. The authors gratefully acknowledge the valuable discussions and insights
provided by the members of the Collaboration on Precision Tests and New Physics (CPTNP).

\end{acknowledgments}

\bibliography{biblio}

\clearpage
\onecolumngrid

\begin{center}
    \textbf{\large Supplemental Material}
\end{center}

\vspace{2em}

\setcounter{equation}{0}
\setcounter{figure}{0}
\setcounter{table}{0}
\setcounter{page}{1}
\setcounter{section}{0}
\makeatletter
\renewcommand{\theequation}{S\arabic{equation}}
\renewcommand{\thefigure}{S\arabic{figure}}
\renewcommand{\thetable}{S\arabic{table}}
\renewcommand{\thesection}{S\arabic{section}}
\makeatother

\twocolumngrid
\section{I. Implementation of \collideranget}

\subsection{Skill-Based Multi-Agent Implementation}
\label{sec:skills}

The implementation used in this work is realized in \textsc{Claude Code}~\cite{claude_code}, i.e. on top of the \textsc{Claude Agent SDK}~\cite{claude_agent_sdk}. This choice provides a practical and user-friendly instantiation of the proposed architecture while allowing the agent to interoperate naturally with existing file, shell, and web-facing tools. At the same time, the workflow logic is not tied to this particular runtime: the central abstraction is a portable \textsc{Agent Skill}~\cite{AgentSkills}, which allows the same high-level design to be reused across agent frameworks.

A skill is organized as a lightweight directory containing a machine-readable \texttt{SKILL.md} file together with optional reference documents and templates. The \texttt{SKILL.md} file specifies the skill name, trigger conditions, expected inputs and outputs, workflow guidance, parameter conventions, and example invocations. A \texttt{references} subdirectory provides condensed domain documents, including syntax rules, software-specific conventions, command references, and annotated examples, while optional \texttt{templates} files provide structured starting points such as skeleton model files. The reference material associated with each skill is based on official software documentation~\cite{Alloul:2013bka,Alwall:2014hca,Conte:2012fm}, systematically curated by the authors on the basis of domain expertise and refined through extensive iterative testing. In this way, a skill acts as a portable and interpretable domain handbook that an agent can read at runtime, rather than as a rigid hard-coded program. Because \textsc{Agent Skills} is now an open standard for packaging agent capabilities and domain knowledge~\cite{AgentSkills}, the implementation can be transferred readily across agent frameworks. As described in the Installation and Usage section, the same skills can also be used directly by other coding agents, including \textsc{Cursor}~\cite{cursor} and \textsc{Codex}~\cite{codex}.

In the current implementation, each \textsc{Agent Skill} is invoked through a dedicated sub-agent, and each sub-agent is responsible for one phenomenological sub-task within the overall workflow. This realizes the multi-agent architecture introduced in the main text in a concrete operational form: rather than asking a single agent to handle the entire pipeline, the system decomposes the user request into specialized stages that can be executed by separate sub-agents with task-specific context.

The current implementation contains five main sub-agents. The \texttt{orchestrator} sub-agent is equipped with the \texttt{pheno-pipeline-orchestrator} skill, which decomposes the natural-language user request into sub-tasks that can be handled independently by the downstream sub-agents. The \texttt{model-generator} sub-agent uses three skills, namely \texttt{feynrules-model-generator}, \texttt{feynrules-validator}, and \texttt{ufo-generator}. Here \texttt{feynrules-model-generator} translates a theoretical Lagrangian written in \LaTeX\ into a \feynrules\ \texttt{.fr} model file; \texttt{feynrules-validator} checks the resulting \texttt{.fr} file for self-consistency through syntax checks, built-in \feynrules\ consistency tests such as Hermiticity, and test loading of the exported UFO model in \madgraph; and \texttt{ufo-generator} converts the validated \texttt{.fr} model into UFO format. Taken together, these skills  allow the \texttt{model-generator} sub-agent to transform a natural-language model specification into a UFO model ready for collider simulation.

The \texttt{collider-simulator} sub-agent uses the \texttt{madgraph-simulator} skill, which writes \madgraph\ scripts and cards from the user's natural-language instructions and executes parton-level simulation, parton showering with \pythia, and detector simulation with \delphes. The \texttt{event-analyzer} sub-agent uses the \texttt{madanalysis-analyzer} skill, which performs event-level analysis with \madanalysis, including cuts, cutflow construction, and histogram extraction; the current implementation is optimized for the \madanalysis\ normal mode. Our preliminary tests with the Large Language Model (LLM) Claude Opus 4.6 further indicate that many expert-mode analysis tasks can already be carried out directly by the LLM through reading \texttt{LHE} or \texttt{LHCO} files, or \texttt{ROOT} files via the \texttt{Scikit-HEP}~\cite{Rodrigues:2020syo} packages \texttt{uproot} and \texttt{awkward}, followed by programmatic implementation of the required selection and cutflow logic. Finally, the \texttt{pheno-analyzer} sub-agent does not currently rely on a dedicated skill; instead, it uses the reasoning capability of the underlying LLM to perform downstream phenomenological inference and visualization, including profile-likelihood analysis, exclusion-limit extraction, parameter scans, and plotting. In addition to these task-specific skills, a shared \texttt{magnus utility skill} provides the operational interface to the \magnus\ platform, including job submission, status monitoring, reruns, output retrieval, and error recovery.

The advantage of this multi-agent design is that each sub-agent can operate with a complete and independent local context, which improves the reliability of sub-task execution. To coordinate information across sub-agents, the system maintains structured intermediate progress records. Upon completion of a task, each sub-agent writes a compact Markdown summary to the \texttt{progress} directory, recording the essential physical assumptions, software-state information, and intermediate outputs needed by later stages. Subsequent sub-agents then read the relevant progress records when continuing the workflow. As discussed in the main text, these summaries are deliberately designed to retain the necessary information while minimizing context overhead. Together, the portable skill abstraction and the structured progress records allow the multi-agent system to translate high-level physical objectives into coherent tool-specific operations across the full workflow.

\subsection{ADK-based Reference Implementation}
\label{sec:adk_impl}

To complement the Claude-Code-based implementation used in this work, we also built a reference system in \textsc{Python} using the \textsc{Google ADK}~\cite{google_adk}. This implementation serves two purposes. First, it provides a concrete and partially reproducible baseline showing how an LLM-based agent can orchestrate the collider-phenomenology workflow through tool use, staged execution, and intermediate-result management. Second, it illustrates the portability of our skill-based architecture by showing that the same high-level design can be instantiated in a different agent framework.

The ADK-based agent is equipped with two classes of tools. \texttt{Filesystem tools} provide access to local files, allowing the agent to read task instructions and reference documents, maintain intermediate artifacts, and write structured outputs for each stage. \texttt{MCP tools} connect the agent to the \magnus platform, enabling environment setup and remote execution of domain-specific software such as \mathematica and \madgraph. Together, these tools allow the agent to traverse the workflow from model construction to simulation and analysis within a unified execution framework. To support reliable execution, the reference implementation uses the same curated domain references as the skill-based system.

\subsection{The \magnus\ Execution Backend}

\magnus~\cite{Magnus} is an open-source platform that turns computing infrastructure into a unified execution backend where both humans and AI agents submit jobs, run containerized toolchains, and crystallize validated workflows into reusable artifacts. It is organized around three layers: an \emph{execution} layer that runs containerized jobs with filesystem isolation and automatic image caching; a \emph{sedimentation} layer in which Blueprints and Skills form a directed knowledge graph that accumulates institutional knowledge; and a \emph{collaboration} layer for shared governance across roles.

The central abstraction is the \emph{Blueprint}: a typed Python function whose signature defines parameters using \texttt{Annotated} type metadata, and whose body defines how a job is submitted. The platform introspects the function to simultaneously generate a web form, validate inputs from the CLI, and expose a programmatic API. This means the same Blueprint can be launched by a researcher through the web UI, from the terminal via \texttt{magnus run <blueprint-id>}, or by an agent through the SDK --- with identical execution semantics and full auditability in all cases. This human-agent symmetry is a deliberate design choice: the platform does not distinguish between a human clicking a button and an agent calling an API.

Blueprints are not static artifacts. Agents can create, execute, evaluate, and refine them, closing the loop between experimentation and sedimentation. A workflow that starts as a one-off experiment can be crystallized into a Blueprint; an agent can later improve it based on new results, guided by the domain knowledge encoded in Skills. In this work, the five \collideranget Blueprints described above were authored by domain experts and are bundled with the \magnus SDK; the self-evolution capability is designed for future autonomous research loops in which the agent iterates on its own toolchain.

\section{II. Details on physics validation and additional benchmarks}

\subsection{Additional Benchmarks}

In addition to the four representative benchmarks presented in the main text, we further tested \collideranget\ on several literature reproductions that broaden the coverage in both physics content and collider environment. 

\begin{figure*}[t]
    \centering
    \subfloat{\includegraphics[width=0.48\textwidth]{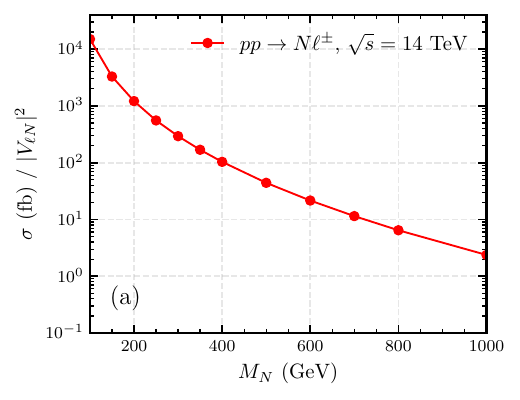}}
    \hfill
    \subfloat{\includegraphics[width=0.48\textwidth]{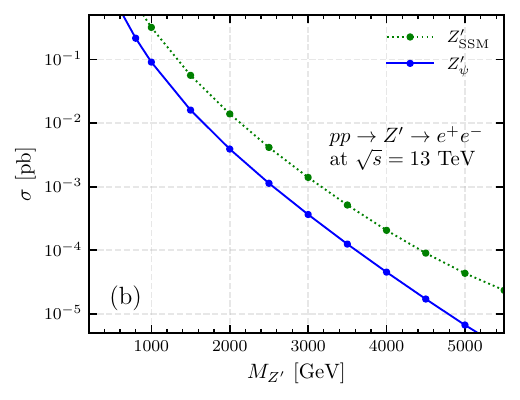}}

    \subfloat{\includegraphics[width=0.48\textwidth]{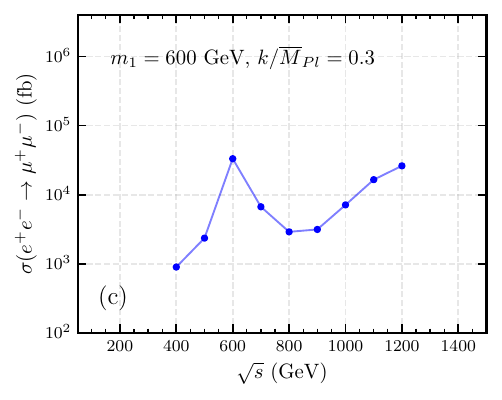}}
    \hfill
    \subfloat{\includegraphics[width=0.48\textwidth]{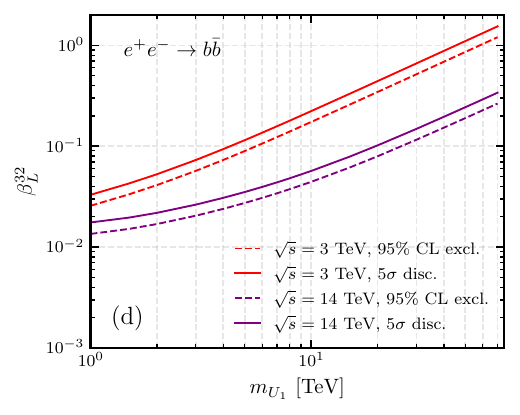}}
    \caption{Additional literature reproductions by \collideranget. (a) Production cross section for $pp \to \mu^\pm N$ at the LHC as a function of $m_N$. (b) Dilepton production cross section for a general $Z^\prime$ benchmark at the $13~\mathrm{TeV}$ LHC as a function of $M_{Z^\prime}$ for the SSM and $Z^\prime_\psi$ scenarios. (c) Cross section for $e^+e^- \to \mu^+\mu^-$ as a function of $\sqrt{s}$ in the Randall--Sundrum model, showing the resonant structure of the KK graviton tower. (d) 95\% CL exclusion and $5\sigma$ discovery contours for the $U_1$ leptoquark benchmark from $\mu^+\mu^- \to b\bar b$ at a muon collider. Panels (a)-(d) reproduce Refs.~\cite{Dev:2013wba,CMS:2021ctt,Davoudiasl:1999tf,Asadi:2021gah}, respectively. See text for details.}
    \label{fig:reproduction_additional}
\end{figure*}

\paragraph{Heavy Majorana Neutrino.} The prompt specified the interaction Lagrangian, the process $pp \to \mu^\pm N$ at $\sqrt{s}=7$, $8$, and $14~\mathrm{TeV}$ LHC, and a parameter scan over $m_N$ with $|V_{\mu N}|=1$. The agent collected cross sections from existing simulation outputs and assembled the corresponding mass dependence across multiple collider energies. As shown in Fig.~\ref{fig:reproduction_additional}(a), the reproduced curves recover the expected hierarchy among the three center-of-mass energies and the decrease of the production rate with increasing $m_N$, in agreement with Ref.~\cite{Dev:2013wba}.

\paragraph{General $Z'$ Benchmark.} The prompt specified a general $Z'$-fermion interaction Lagrangian, the dilepton process $pp \to Z' \to \mu^+\mu^-$, and two benchmark coupling assignments corresponding to the Sequential SM (SSM)~\cite{Altarelli:1989ff, Accomando:2010fz} and an $E_6$-inspired $Z'_\psi$ scenario~\cite{Hewett:1988xc, Leike:1998wr}. The agent was instructed to scan the resonance mass and extract the cross section for each benchmark. As shown in Fig.~\ref{fig:reproduction_additional}(b), the resulting cross-section curves reproduce the expected mass dependence and the characteristic normalization difference between the SSM and $Z'_\psi$ scenarios, consistent with Ref.~\cite{CMS:2021ctt}.

\paragraph{KK Graviton at a Lepton Collider.} We considered the Randall-Sundrum model~\cite{Randall:1999ee} in the process $e^+e^- \to \mu^+\mu^-$, for which the prompt specified the graviton interaction Lagrangian in terms of the SM energy-momentum tensor, the spin-2 Kaluza-Klein (KK) graviton spectrum, and a scan over the collider energy $\sqrt{s}$. Unlike the hadron-collider benchmarks above, this case probes a lepton-collider environment together with higher-rank tensor interactions and multiple massive resonances from the KK tower. As shown in Fig.~\ref{fig:reproduction_additional}(c), the agent successfully extracted the cross section across the scanned energy points and reproduced the resonant lineshape of the KK tower of gravitons, in agreement with Ref.~\cite{Davoudiasl:1999tf}.

\paragraph{$U_1$ Leptoquark at a Muon Collider.} We further considered the $U_1$ leptoquark scenario at a muon collider, for which the prompt specified the interaction Lagrangian, the Drell--Yan process $\mu^+\mu^- \to b\bar b$, and the flavor structure in which only $\beta_L^{32}$ is nonzero. The task was to derive the 95\% CL exclusion and $5\sigma$ discovery contours in the $(m_{U_1},\,\beta_L^{32})$ plane for $\sqrt{s}=3$ and $14~\mathrm{TeV}$. As shown in Fig.~\ref{fig:reproduction_additional}(d), the agent successfully reproduced the reach contours, consistent with Ref.~\cite{Asadi:2021gah}. This example extends the validation beyond the LHC benchmarks in the main text and demonstrates that the framework can also handle interference-driven observables and sensitivity projections at future muon colliders.

\subsection{Details on Physics Validation}

For the physics-validation benchmarks, we used \textsc{Claude Code} with the skill-based multi-subagent framework, running on the Claude Opus 4.6 model. For each benchmark prompt, we performed three independent reproduction runs.

To eliminate shortcuts from persistent memory or previously generated results, we designed sandboxed evaluation pipeline based on \texttt{bwrap}. In each run, the agent was given only a clean workspace initialized with a single \texttt{prompt.md} file, together with the minimal runtime dependencies required for execution. This setup ensured that every reproduction started from the same isolated initial state and that the comparison across runs reflected the agent's actual problem-solving capability rather than reuse of prior context.

Across the three independent runs for each benchmark prompt (see the Example Natural-Language Prompts section), \collideranget\ often produced outputs that were consistent with the target benchmark results, although occasional failed reproductions were also observed. Representative successful reproductions are shown in Fig.~\ref{fig:reproduction} of the main text and Fig.~\ref{fig:reproduction_additional} of this Supplemental Material.

\section{III. Installation and Usage}
\label{sec:installation}

Full installation instructions and up-to-date usage guides are maintained at \url{https://github.com/HET-AGI/ColliderAgent}; below we summarize the essential steps. \collideranget delegates all computational tasks to \magnus, which manages the underlying HEP toolchain through two pre-built container images: \texttt{collider} (\madgraph, \pythia, \delphes, \madanalysis) and \texttt{mma-het} (\mathematica, \feynrules), each pulled on demand the first time a blueprint requires it.

\paragraph{Prerequisites.}
The following tools must be installed and available on the host system:
\begin{itemize}
\item A coding agent --- \textsc{Claude Code}~\cite{claude_code} is recommended for full multi-agent support; \textsc{Codex}~\cite{codex}, \textsc{Cursor}~\cite{cursor}, and others work in skills-only mode
\item Python~$\geq$~3.10
\item Docker (with the daemon running)
\item Git
\item uv (Python package manager; \texttt{pip install uv})
\item Node.js --- optional but recommended; enables the Web UI for visual job monitoring
\end{itemize}
No HEP-specific software needs to be installed on the host. On Windows, a Bash-compatible shell (e.g., WSL or Git Bash) is needed to run the shell commands below.

\paragraph{Step 1: Clone and install.}
Run the following commands in a terminal:
\begin{lstlisting}[basicstyle=\ttfamily\footnotesize,columns=flexible,xleftmargin=4pt]
git clone https://github.com/HET-AGI/ColliderAgent.git
cd ColliderAgent
pip install -e .
\end{lstlisting}
The \texttt{magnus-sdk} dependency is installed automatically, providing the \texttt{magnus} CLI and the server-side blueprints described in the main text.

\paragraph{Step 2: Launch the local backend.}
Run the following command in a terminal:
\begin{lstlisting}[basicstyle=\ttfamily\footnotesize,columns=flexible,xleftmargin=4pt]
magnus local start
\end{lstlisting}
This command fetches the \magnus source repository (if not already present), installs backend dependencies, starts the backend server (port~8017), creates a local database and user account, and registers all bundled blueprints. If Node.js is installed, a web interface is also launched at \texttt{http://localhost:3011}. To verify that the backend is running correctly:
\begin{lstlisting}[basicstyle=\ttfamily\footnotesize,columns=flexible,xleftmargin=4pt]
magnus run hello-world
\end{lstlisting}
A successful run prints \texttt{Hello from Magnus!} after pulling the required container image.

\paragraph{Step 3: Install skills and sub-agents.}
\collideranget is organized into reusable \emph{skills} (domain knowledge and templates for each HEP tool) and \emph{sub-agents} (specialized agents that orchestrate subsets of skills). Both must be copied to directories that the coding agent reads at startup. For \textsc{Claude Code}~\cite{claude_code}:
\begin{lstlisting}[basicstyle=\ttfamily\footnotesize,columns=flexible,xleftmargin=4pt]
cp -r src/skills ~/.claude/skills
cp -r src/agents ~/.claude/agents
\end{lstlisting}
The full multi-agent architecture (skills + sub-agents) currently requires \textsc{Claude Code}. Other coding agents such as \textsc{Codex}~\cite{codex} and \textsc{Cursor}~\cite{cursor} can run a skills-only version of \collideranget by copying \texttt{src/skills/} to the agent's global skills directory; see the repository README for agent-specific paths.

\paragraph{Step 4: Run the \collideranget hello-world prompt.}
The following prompt serves as a minimal end-to-end test of the full pipeline:
\begin{quote}
\itshape
Plot the dilepton invariant mass distribution for parton-level $p p \to \ell^+ \ell^-$ process at the 14\,TeV LHC in the SM.
\end{quote}
Using \textsc{Claude Code} as the coding agent:
\begin{lstlisting}[basicstyle=\ttfamily\footnotesize,columns=flexible,xleftmargin=4pt]
claude -p "Plot the dilepton invariant mass distribution
  for parton-level pp -> l+l- process at the
  14 TeV LHC in the SM."
\end{lstlisting}
Because this prompt uses only the Standard Model, it requires only the \texttt{collider} image ($\sim$940\,MB), which is pulled automatically on the first invocation and cached for subsequent runs. The three blueprints involved are \texttt{madgraph-compile}, \texttt{madgraph-launch}, and \texttt{madanalysis-process}; the \texttt{mma-het} image is not needed. Intermediate artifacts are stored in the working directory for inspection and iterative refinement.

\paragraph{Step 5: Activate the Wolfram Engine license (one-time, for custom BSM models).}
The two \feynrules-based blueprints (\texttt{validate-feynrules} and \texttt{generate-ufo}) require a Wolfram Engine license. Because the license is tied to the machine identity of the \emph{container} rather than the host, activation must be performed inside the container itself. First, register a free Wolfram ID at \url{https://wolfram.com/engine/free-license}. Then run:
\begin{lstlisting}[basicstyle=\ttfamily\footnotesize,columns=flexible,xleftmargin=4pt]
mkdir -p ~/.wolfram-container-license
IMAGE=git.pku.edu.cn/2200011523/mma-het:latest
MOUNT=~/.wolfram-container-license
docker run -it --rm \
  -v $MOUNT:/root/.WolframEngine/Licensing \
  $IMAGE wolframscript
\end{lstlisting}
Follow the interactive prompts to enter your Wolfram ID and password. The license file (\texttt{mathpass}) is written to the mounted host directory \texttt{\textasciitilde/.wolfram-container-license/}, which all subsequent \feynrules\ blueprint runs mount automatically.

\paragraph{Cloud deployment.}
For deployment on HPC clusters with \textsc{Slurm}, the command \texttt{magnus login} stores credentials and routes all subsequent commands to the remote backend; agent skills, CLI syntax, and job artifacts remain identical in both modes. All benchmarks in this work use the cloud deployment.

\subsection{Example Natural-Language Prompts}

\noindent Below we provide the user prompts corresponding to the reproductions shown in Fig.~\ref{fig:reproduction}(a), (b), and (d) in the main text. For the benchmark evaluation, these prompts are written with a relatively high level of detail in order to maximize the reproduction success rate and make the task specification fully explicit.

In practical use of \collideranget, however, many of these details are not essential. For example, in the prompt for Fig.~\ref{fig:reproduction}(b), one could omit the explicit definitions of the SM field-strength tensors $B_{\mu\nu}$ and $W^a_{\mu\nu}$. Likewise, in the prompt for Fig.~\ref{fig:reproduction}(d), the experimental data could instead be retrieved by the agent from the web, e.g. by prompting it to search for and use the ATLAS data in a specific paper such as arXiv:1801.06992. Similarly, one could provide only the arXiv identifiers of the relevant ATLAS and CMS analyses and ask the agent to read the papers and identify the corresponding selection cuts. These examples illustrate benchmark specifications rather than the minimum information required in routine use.

The complete prompts for all literature reproductions shown in Fig.~\ref{fig:reproduction} of the main text and Fig.~\ref{fig:reproduction_additional} of the Supplemental Material are available in the \collideranget\ GitHub repository. We also note that, in order to best demonstrate the capability of \collideranget, the concrete reproduction workflow in these prompts may differ in some technical details with the procedure adopted in the original paper, even when the final physical result being reproduced is the same.

\onecolumngrid


\vspace{2em}

\begin{promptblocktex}[User Prompt Input for reproducing the result in Ref.\cite{Buonocore:2020erb}, i.e., Fig.\ref{fig:reproduction}(a)]

\textbf{1. Target}

\medskip

Considering the scalar leptoquark model described in this document, plot the
$m_{ej}$ distribution for the signal process.

\medskip
\hrule
\medskip

\textbf{2. Scalar Leptoquark Model}

\medskip

\textbf{2.1 Lagrangian}

\[
\mathcal{L} = \lambda_{eu}\,\text{LQ}_{eu}\, \bar e_R u_R^c + \text{h.c.}
\]
where $\psi^c \equiv C \bar\psi^T$ with $C = i \gamma^2 \gamma^0$ denotes the charge conjugation of $\psi$ field. Note that
the leptoquark $\text{LQ}_{eu}$ couples to a lepton and a quark (no anti-particles). Here,
\begin{itemize}
  \item $\text{LQ}_{eu}$ is a scalar leptoquark. It is an $SU(2)_L$ singlet under the SM gauge group, carrying color triplet and electric charge $Q = -1/3$.
  \item $\lambda_{eu}$ is the Yukawa coupling (real).
\end{itemize}

\textbf{2.2 Parameters}

The benchmark point for the signal:
\begin{itemize}
  \item mass of LQ: $M_\text{LQ} = 3000$ GeV
  \item total width of LQ: $\Gamma_\text{LQ} = 60$ GeV
  \item $\lambda_{eu} = 1$
\end{itemize}

\medskip
\hrule
\medskip

\textbf{3. Collider Simulation}

\medskip

\textbf{3.1 Process}

\[
p\,p \to \text{LQ} \to e\,j
\]

This is resonant single leptoquark production via lepton-quark fusion. Here, the underlying process is a lepton ($e$) from one proton PDF and a quark ($u$)
from the other proton fusion to produce the LQ, which then decays back to $e + j$.
This requires the LUXlep PDF, which provides lepton parton distribution functions
inside the proton.

\textbf{3.2 Collider simulation settings}

\begin{itemize}
  \item Collider: 13 TeV LHC
  \item Event number: 100000
  \item Parton shower: Pythia8
  \item Detector simulation: Delphes with ATLAS card (anti-$k_T$ jets with $R = 0.4$)
  \item PDF: LUXlep; redefine the proton content to include leptons and the photon
  \item Generation-level cuts: $p_T(\ell, j) > 500$ GeV, $|\eta| < 2.5$
  \item Store the generated events in LHCO format
\end{itemize}

\textbf{3.3 Pythia8 lepton-to-photon workaround}

Pythia8 cannot backward-evolve leptons from proton PDFs. The correct steps are:
\begin{enumerate}
  \item After MadGraph generates the LHE file, replace all initial-state leptons with photons in the LHE file.
  \item Disable Pythia8's built-in event validity checks (charge/momentum conservation) so it accepts and showers the manually modified LHE file without rejecting it.
  \item Perform shower and detector simulation.
\end{enumerate}

\medskip
\hrule
\medskip

\textbf{4. Numerical Analysis}

\medskip

\textbf{4.1 Event selection}

Read the reconstructed events from the Delphes output and apply the following selection cuts:
\begin{enumerate}
  \item Electron: $p_T > 500$ GeV, $|\eta| < 2.5$
  \item Jet: $p_T > 500$ GeV, $|\eta| < 2.5$ (anti-$k_T$, $R = 0.4$)
  \item Missing transverse energy: $E_T^\text{miss} < 50$ GeV
  \item Lepton veto: veto events with additional leptons ($|\eta| < 2.5$, $p_{T,\ell} > 7$ GeV)
  \item Jet veto: veto events with additional subleading jets ($|\eta| < 2.5$, $p_{T,j} > 30$ GeV)
\end{enumerate}

\textbf{4.2 Signal histogram}

Compute the invariant mass $m_{ej}$ of the leading electron and leading jet for events passing all cuts.

\begin{itemize}
  \item Bin the $m_{ej}$ distribution in 100 GeV bins from 0 to 5000 GeV.
  \item Weight each event by $
  w = \sigma \times \mathcal{L} / N_\text{gen}$, 
  where $\sigma$ is the cross section, $\mathcal{L} = 100\;\text{fb}^{-1}$, and $N_\text{gen}$ is the total number of generated events.
\end{itemize}

\medskip
\hrule
\medskip

\textbf{5. Plot Figure}

Plot the signal $m_{ej}$ distribution with solid black line.

\end{promptblocktex}
\begin{promptblocktex}[User Prompt Input for reproducing the result in Ref.\cite{Brivio:2017ije}, i.e., Fig.\ref{fig:reproduction}(b).]

\textbf{1. Target}

\medskip

Considering the ALP EFT in this document, plot the normalized
$E_T^{\text{miss}}$ distribution for
$pp \to a\,W^\pm\gamma$ ($W^\pm \to \ell^\pm\nu$)
at $\sqrt{s} = 13$ TeV LHC.

\medskip
\hrule
\medskip

\textbf{2. ALP EFT}

\medskip

\textbf{2.1 Lagrangian}

The ALP bosonic EFT Lagrangian reads:

\[
\delta\mathcal{L}_a^{\text{bosonic}}
=
c_{\tilde{W}} \mathcal{A}_{\tilde{W}}
+
c_{\tilde{B}} \mathcal{A}_{\tilde{B}}
\]

where the operators are

\[
\mathcal{A}_{\tilde{B}}
=
-B_{\mu\nu} \tilde{B}^{\mu\nu} \frac{a}{f_a},
\qquad
\mathcal{A}_{\tilde{W}}
=
-W^a_{\mu\nu} \tilde{W}^{a\mu\nu} \frac{a}{f_a}.
\]

Here,

\begin{itemize}
\item $a$ is the ALP (axion-like particle), a pseudo-scalar singlet, with mass $m_a$.
\item $f_a$ is the ALP decay constant (dimension of mass).
\item $B_{\mu\nu} = \partial_\mu B_\nu - \partial_\nu B_\mu$ is the U(1)$_Y$ hypercharge field strength tensor.
\item $W^a_{\mu\nu} =
\partial_\mu W^a_\nu -
\partial_\nu W^a_\mu +
g \epsilon^{abc} W^b_\mu W^c_\nu$
is the SU(2)$_L$ weak isospin field strength tensor ($a = 1,2,3$).
\item $\tilde{B}^{\mu\nu}
=
\frac{1}{2}\epsilon^{\mu\nu\rho\sigma}B_{\rho\sigma}$,
$\tilde{W}^{a\mu\nu}
=
\frac{1}{2}\epsilon^{\mu\nu\rho\sigma}W^a_{\rho\sigma}$
are the dual field strength tensors.
\item $c_{\tilde{W}}, c_{\tilde{B}}$ are dimensionless Wilson coefficients.
They satisfy the relation
\[
c_{\tilde{B}}
=
-\tan^2\theta_W \cdot c_{\tilde{W}}
\]
to enforce $g_{a\gamma\gamma} = 0$,
where $\theta_W$ is the weak mixing angle.
\item In the collider simulation,
$f_a = 1000$ GeV and $m_a = 0.001$ GeV are chosen.
So the free parameter is $c_{\tilde{W}}$.
\end{itemize}

\medskip
\hrule
\medskip

\textbf{3. Collider Simulation}

\medskip

\textbf{3.1 Process}

\[
p\,p \to a\,W^\pm\,\gamma,
\qquad
W^\pm \to \ell^\pm \nu
\]

\textbf{3.2 Collider Simulation Settings}

\begin{itemize}
\item Collider: 13 TeV LHC
\item Event number: 500,000
\item Analysis level: Parton-level (no parton shower or detector simulation)
\item PDF: nn23lo1
\item $f_a = 1000$ GeV, $m_a = 0.001$ GeV, and $c_{\tilde{W}} = 1$
\end{itemize}

\medskip
\hrule
\medskip

\textbf{4. Numerical Analysis}

\medskip

\textbf{4.1 Event Selection}

Read the LHE events and apply the following selection cuts:

\begin{itemize}
\item photon: $p_T > 20$ GeV, $\eta < 2.5$
\item lepton: $p_T > 20$ GeV, $\eta < 2.5$
\end{itemize}

\textbf{4.2 Histogram and Normalization}

\begin{itemize}
\item Histogram: $E_T^{\text{miss}}
=
\left|
\vec{p}_T^{\,a} + \vec{p}_T^{\,\nu}
\right|$, the vector sum of the transverse momenta of all invisible particles
(ALP + neutrino).

\item Binning: 50 bins, 0--1000 GeV
\end{itemize}

\textbf{4.3 Plot Figure}

Plot the histogram. The height of the histogram is the normalized events,
which are defined as the ratio of the number of events in the bin to the
total number of events.

\end{promptblocktex}
\begin{promptblocktex}[User Prompt Input for reproducing the result in Ref.\cite{Greljo:2018tzh}, i.e., Fig.\ref{fig:reproduction}(d).]

\textbf{1. Target}

\medskip

Considering the $U_1$ Leptoquark Model described in this document,
plot the $2\sigma$ exclusion contour for the $U_1$ leptoquark
in the $(\sqrt{|g_c^* g_b|},\, M_{U_1})$ plane,
with some other lines and bands.

\medskip
\hrule
\medskip

\textbf{2. $U_1$ Leptoquark Model}

In this section, the $U_1$ Leptoquark Model is introduced.

\medskip

\textbf{2.1 Lagrangian}

\[
\mathcal{L}_{U_1}
=
-(D_\mu U_{1\nu}-D_\nu U_{1\mu})^\dagger
(D^\mu U_1^\nu-D^\nu U_1^\mu)
+
M_{U_1}^2\,U_{1\mu}^\dagger U_1^\mu
+
\bigl[
g_c(\bar c\gamma^\mu P_L\nu_\tau)U_{1\mu}^\dagger
+
g_b(\bar b\gamma^\mu P_L\tau)U_{1\mu}^\dagger
+
\text{h.c.}
\bigr]
\]

Here,

\begin{itemize}
\item $U_1$ is a vector leptoquark beyond the SM.
It is a new gauge boson carrying color triplet,
$SU(2)_L$ singlet, and electric charge $Q=2/3$.
\item $D_\mu$ is the covariant derivative of the SM gauge fields.
\end{itemize}

\medskip

\textbf{2.2 Parameters}

The free parameters are:

\begin{itemize}
\item $M_{U_1}$ is the mass of $U_1$.
\item $g_c$ and $g_b$ are the couplings of $U_1$ to $c$ and $b$ quarks,
respectively. Both of them are real.
\end{itemize}

\medskip
\hrule
\medskip

\textbf{3. Collider Simulation}

\medskip

\textbf{3.1 Process}

\[
pp \to \tau \nu
\]

which is mediated by the $U_1$ vector leptoquark.
Since the $U_1$ leptoquark couples to $b$ and $c$ quarks,
the initial-state $b$ and $c$ partons from the proton PDF must be included.

\medskip

\textbf{3.2 Collider simulation settings}

We have two runs. For each run,

\begin{itemize}
\item collider: 13 TeV LHC
\item event number: 10000
\item parton shower: Pythia8
\item perform mass scan for $M_{U_1}$ masses:
750, 1000, 1250, 1500, 2000, 2500, 3000, 4000, 5000 GeV
\item output format for reconstructed events: LHCO
\item comment: decay width of LQ is not relevant,
since only the $t$-channel contribution exists.
\end{itemize}

Run 1 (with ATLAS detector simulation):

\begin{itemize}
\item detector simulation: Delphes ATLAS card
\end{itemize}

Run 2 (with CMS detector simulation):

\begin{itemize}
\item detector simulation: Delphes CMS card
\end{itemize}

\medskip
\hrule
\medskip

\textbf{3 Numerical Analysis}

\medskip

\textbf{3.1 Experimental data}

Binned $m_T$ distribution from ATLAS and CMS are given below.

\medskip

\textbf{ATLAS}

\begin{itemize}
\item Stored as \texttt{analysis/hepdata/table1.yaml}
\item 22 bins from 250 to 3200 GeV (log-spaced)
\item Columns: observed events $n_i$, SM background $b_i$,
symmetric error $\delta b_i$
\end{itemize}

\medskip

\textbf{CMS}

The experimental data is given below:

\medskip

\begin{center}
\begin{tabular}{cccc}
$m_T$ bin (GeV) & $n_{\text{obs}}$ & $b_{\text{SM}}$ & $\delta b$ \\
\hline
320–500 & 1203 & 1243 & 160 \\
500–1000 & 452 & 485 & 77 \\
1000–3200 & 15 & 23.4 & 6.2
\end{tabular}
\end{center}
\medskip

\textbf{3.2 Simulated signal events}

In this section, the expected signal events in each bin are calculated.

\medskip

\textbf{step 1: event selection}

Read the reconstructed events from the simulation output
and apply experiment-specific selections.

\medskip

\textbf{ATLAS selection}

\begin{itemize}
\item Lepton veto: no electrons or muons
\item $\ge 1$ hadronic tau with $p_T>80$ GeV, $|\eta|<2.3$
\item $E_T^{\text{miss}}>150$ GeV
\item $m_T>250$ GeV, with $m_T=\sqrt{2p_T^\tau E_T^{\text{miss}}(1-\cos\Delta\phi)}$
\end{itemize}

\medskip

\textbf{CMS selection}

\begin{itemize}
\item Lepton veto: no electrons or muons
\item $\ge 1$ hadronic tau with $p_T>80$ GeV, $|\eta|<2.1$
\item $E_T^{\text{miss}}>200$ GeV
\item $0.7 < p_T^\tau/E_T^{\text{miss}} < 1.3$
\item $\Delta\phi(\tau,E_T^{\text{miss}}) > 2.4$
\item $m_T>320$ GeV
\end{itemize}

\medskip

\textbf{step 2: signal template construction}

For events passing selection,
histogram $m_T$ into ATLAS or CMS bins.

\[
s_i^{(g=1)}
=
\frac{N_i^{\text{pass}}}{N_{\text{gen}}}
\times
\sigma(g=1)
\times
\mathcal{L}
\]

where

\begin{itemize}
\item $N_{\text{gen}}$ is the number of reconstructed events
\item $\sigma(g=1)$ is the cross section at coupling $g=1$
\item $\mathcal{L}$ is the luminosity
(36.1 fb$^{-1}$ ATLAS, 35.9 fb$^{-1}$ CMS)
\end{itemize}

Signal scaling: $s_i(g)=g^4 s_i^{(g=1)}$

\medskip

\textbf{3.3 Profile likelihood analysis}

For each bin,

\[
-\ln L_i(\theta_i)
=
-\bigl[n_i\ln\mu_i-\mu_i\bigr]
+
\frac12
\frac{\theta_i^2}{(\delta_i/b_i)^2}
\]

where

\begin{itemize}
\item $n_i$ observed events
\item $b_i$ expected SM background
\item $\delta_i$ systematic uncertainty
\item $\theta_i$ nuisance parameter
\item $s_i(g)=g^4 s_i^{(g=1)}$
\item $\mu_i=b_i(1+\theta_i)+s_i$
\end{itemize}

\medskip

\textbf{Procedure to find the $2\sigma$ exclusion contour}

\textbf{step 1 profiling:} Minimize $-\ln L_i(\theta_i)$ numerically with constraint $\mu_i>0$.

\medskip

\textbf{step 2 combine ATLAS + CMS Likelihood:} $
\ln\mathcal{L}_{\text{comb}}
=
\ln\mathcal{L}_{\text{ATLAS}}
+
\ln\mathcal{L}_{\text{CMS}}
$

\medskip

\textbf{step 3 extract exclusion region:}

\begin{enumerate}
\item Find best-fit $\hat g$
\item Find the exclusion curve by considering $g_{\text{excl}}$ satisfying
$
-2\bigl[\ln\mathcal{L}(g_{\text{excl}})-\ln\mathcal{L}(\hat g)\bigr]=4
$
\end{enumerate}

\medskip
\hrule
\medskip

\textbf{3.4 Plot Figure}

Plot the exclusion curve obtained from section 3.3 and show the excluded region above the curve with gray shading.

\end{promptblocktex}

\end{document}